\documentclass[prl,twocolumn,showpacs]{revtex4}
\usepackage{amsmath}
\usepackage[dvips,xdvi]{graphicx}
\usepackage{amssymb}
\usepackage{epsfig}

\begin{document}
\title{Observation of Strong Quantum Depletion in a Gaseous Bose-Einstein Condensate}
\author{K. Xu\footnote{Contact Info: kwxu@mit.edu}, Y. Liu, D.E. Miller, J.K. Chin, W. Setiawan and W. Ketterle\footnote{Website: cua.mit.edu/ketterle\_group}}

\affiliation{Department of Physics, MIT-Harvard Center for Ultracold Atoms, and Research Laboratory of Electronics, MIT, Cambridge, MA 02139}
\date{\today}

\begin{abstract}
We studied quantum depletion in a gaseous Bose-Einstein condensate. An optical lattice enhanced the atomic interactions and modified the dispersion relation resulting in strong quantum depletion. The depleted fraction was directly observed as a diffuse background in the time-of-flight images. Bogoliubov theory provides a semi-quantitative description for our observations of depleted fractions in excess of $50 \%$.
\end{abstract}


\maketitle

The advent of Bose-Einstein condensates (BEC) in 1995 extended the study of quantum fluids from liquid helium to superfluid gases with a 100 million times lower density. These gaseous condensates featured relatively weak interactions and a condensate fraction close to 100\%, in contrast to liquid helium where the condensate fraction is only 10\% \cite{helium}. As a result, the gaseous
condensates could be quantitatively described by a single macroscopic wave function shared by all atoms which is the solution of a non-linear Schrodinger equation. This equation provided a mean-field description of collective excitations, hydrodynamic expansion, vortices and sound propagation \cite{dalf99}.

The fraction of the many-body wave function which cannot be represented by the macroscopic wave function is called the quantum depletion. In a homogenous BEC, it consists of admixtures of
higher momentum states into the ground state of the system. The fraction of the quantum depletion $\eta_0$ can be calculated through Bogoliubov theory: $\eta_0 = 1.505\sqrt{\rho a_s^3}$ where $\rho$ is the
atomic density and $a_s$ is the s-wave scattering length \cite{bog}. For $^{23}$Na at a typical density of
$10^{14}$~cm$^{-3}$, the quantum depletion is $0.2\%$.

For the last decade, it has been a major goal of the field to map out the transition from gaseous condensates to liquid helium. Beyond-mean-field effects of a few percent were identified in the temperature dependence of collective excitations in a condensate \cite{kurn,giorgini}. The quantum depletion increases for higher densities -- however, at densities approaching $10^{15}$~cm$^{-3}$
the lifetime of the gas is dramatically shortened by three-body collisions. Attempts to increase the scattering length near a Feshbach resonance were also limited by losses \cite{stenger,Rb85}. Recently, several studies of strongly interacting molecular condensates were performed \cite{Innsbruck, Li2, Bourdel}. In lower dimension systems, the effect of interactions is enhanced. Strongly correlated systems, which are no longer superfluid, were observed in 1D systems \cite{mainz1,weiss}, and in optical lattices \cite{orzel,greiner2}. Quantum depletion in 1D was studied in \cite{kohl1,kohl2,stringari2}, where condensation and quantum depletion are finite-size effects and disappear in the thermodynamic limit. The transition between a three-dimensional quantum gas and a quantum liquid has been largely unexplored.

In this Letter, we report the first quantitative study of strong quantum depletion in a superfluid gas. Exposing atoms to an optical lattice enhances quantum depletion in two ways. First, the lattice increases the local atomic density, which enhances the interactions (by up to an order of magnitude in our experiment), ultimately limited by inelastic collisional losses. The second effect of the lattice is to modify the dispersion relation $T(k)$, which is simply $T(k)=\hbar^2 k^2/2m$ for free atoms. The effect of a weak lattice can be accounted for by an increased effective mass. For a deep lattice, when the width of the first band becomes comparable to or smaller than the interaction energy, the full dispersion relation is required for a quantitative description.

In addition to enhancing the quantum depletion, an optical lattice also enables its direct observation in time of flight.  For a harmonic trap, the quantum depletion cannot be observed during ballistic expansion in the typical Thomas-Fermi regime. Because the mean-field energy (divided by $\hbar$) is much greater than the trap frequency, the cloud remains locally adiabatic during the expansion. The condensate at high density transforms adiabatically into a condensate at low density with diminishing quantum depletion. Therefore, the true momentum distribution of the trapped condensate including quantum depletion and, for the same reason, phonon excitations can only be observed by \textit{in situ} techniques such as Bragg spectroscopy \cite{vogels, stringari}. In an optical lattice, the confinement frequency at each lattice site far exceeds the interaction energy, and the time-of-flight images are essentially a snapshot of the frozen-in momentum distribution at the time of the lattice switch-off, thus allowing for a direct observation of the quantum depletion.

The experiment setup is similar to that of our previous work \cite{kwxu}: A $^{23}$Na BEC containing up to $5\times10^5$ atoms in the $\left | F=1, m_F = -1 \right \rangle$ state was loaded into a crossed optical dipole trap. The number of condensed atoms was controlled through three-body decay in a compressed trap, after which the trap was relaxed to allow further evaporation and re-thermalization. A vertical magnetic field gradient was applied to compensate for gravity and avoid sagging in the weak trap. The final trap frequencies were $\omega_{x,y,z} = 2 \pi \times 60, 60, 85$~Hz. The mean Thomas-Fermi radius was $\sim 12~\mu$m for $1.7\times10^5$ atoms.

The lattice beams were derived from the same single-mode infra-red laser at 1064~nm used for the crossed optical dipole trap. All five beams were frequency-shifted by at least 20~MHz with respect
to each other via acousto-optical modulator to eliminate cross interference between different beams. The three lattice beams had a $1/e^2$-waist of $\sim 90$~$\mu$m at the condensate, and were retro-reflected to form standing waves. The two horizontal beams were orthogonal to each other, while the third beam was tilted $\sim 20^{\circ}$ with respect to the vertical axis due to limited optical access. One and two dimensional lattices were formed using either one or both of the horizontal beams. The trap parameters were chosen such that during the ramping of the optical lattice potential, the overall Thomas-Fermi radii remained approximately constant in order to minimize intra-band excitations. All the
measurements were performed at a peak lattice site occupancy number $\sim7$, as determined by a tradeoff between small three-body losses and good signal-to-noise ratio.

\begin{figure}[htbp]
\includegraphics[width=80mm]{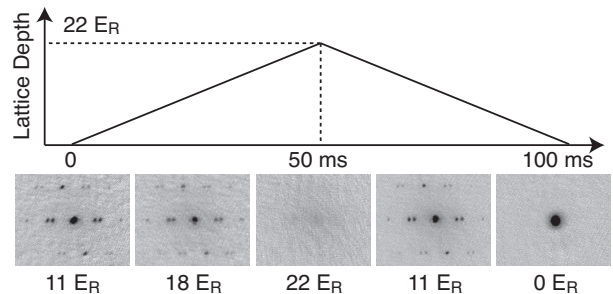}
\caption{Interference patterns in time of flight: The ramping sequence was interrupted as the lattice is ramped up ($11, 18, 22~E_{R}$) and back down ($11, 0~E_{R}$). The time of flight was 10~ms; the field of view is $861~\mu$m$\times 1075~\mu$m.} \label{mott}
\end{figure}

The optical lattice was linearly ramped up to a peak potential of 22$~\pm~2$ $E_{R}$ in time $\tau_{\mathrm{ramp}}$, and then linearly ramped back down at the same speed. This ramp sequence was interrupted at various times by a sudden switch-off of all lattice and trapping potentials ($< 1\mu$s). Absorption images were taken after 10~ms time of flight, reflecting the momentum distribution of the system at the instant of release (Fig.~\ref{mott}). Based on the number of atoms remaining in the condensate after the full ramping sequence ($\gtrsim 80\%$), we concluded that $\tau_{\mathrm{ramp}}\gtrsim 1$~ms satisfies the intra-band adiabaticity condition. In the following discussion, all measurements were performed for $\tau_{\mathrm{ramp}} = 50$~ms.

The loss and revival of the interference contrast, as illustrated in Fig.~\ref{mott}, has been associated with the quantum phase transition from a superfluid to a Mott-insulator \cite{orzel,greiner2}. The presence of sharp interference peaks indicates coherence and superfluid behavior, whereas the presence of a single broad peak indicates the insulating phase. However, as we show in this Letter, even before the lattice depth exceeds the critical value for the phase transition, a diffuse background gradually emerges as a result of quantum depletion. The interference peaks represent the population in the zero quasi-momentum state, and the diffuse background represents the population in the rest
of the Brillouin zone. We only account for the lowest energy band as the population in higher bands remains negligible for our parameters.

In the time-of-flight images, we masked off the sharp interference peaks and performed a two-dimensional gaussian fit on the diffuse background peak. After the lattice was fully ramped down, most of the atoms remained in the condensed fraction while a small fraction (up to $20\%$) were heated and distributed across the first Brillouin zone likely due to the technical noise and imperfect adiabaticity of the ramp. A linear interpolation was used to subtract this small heating contribution (up to $10\%$ at the maximum lattice depth) and obtain the quantum depletion $N_{\mathrm{qd}}/N$, where $N_{\mathrm{qd}}$ is the number of atoms in the diffuse background peak (quantum depletion) and $N$ the total number.

\begin{figure}[htbp]
\includegraphics[width=80mm]{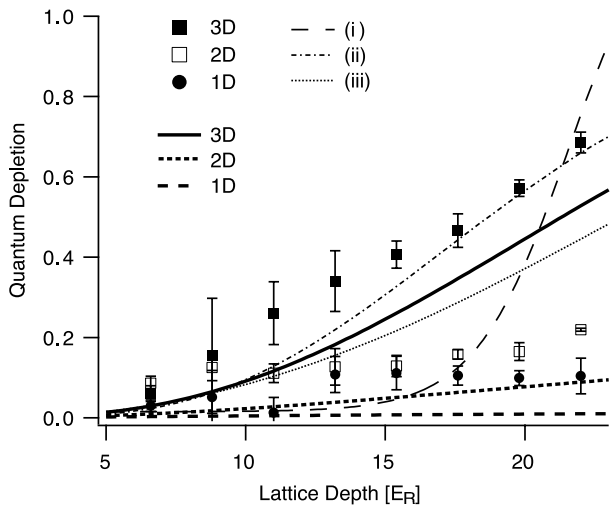}
\caption{Quantum depletion of a $^{23}$Na BEC confined in a one, two and three dimensional optical lattice: the data points with statistical error bars are measured $N_{\mathrm{qd}}/N$; the three thick curves are the theoretical calculation of $N_{\mathrm{qd}}/N$ using Bogoliubov theory and local density approximation. For comparison, also shown are (thin curves): (i) the (smoothed out) Mott-insulator fraction $N_{\mathrm{MI}}/N$ based on a mean-field theory; (ii) the calculated quantum depletion for a \emph{homogeneous} system of per-site occupancy number $n=1$ and (iii) $n=7$.} \label{123d}
\end{figure}

We performed this measurement for one, two and three dimensional optical lattices, and the main results are shown in Fig.~\ref{123d}. The quantum depletion became significant for lattice depths of $\gtrsim 10~E_R$ for a three dimensional lattice ($E_{R}=\hbar^2 k_{\mathrm{latt}}^2/2m$, where $k_{\mathrm{latt}}=2\pi/\lambda_{\mathrm{latt}}$ is the lattice wave number). Note that the Mott-insulator transition starts to occur only at lattice depths $\gtrsim 16~E_R$ (see below). $N_{\mathrm{qd}}/N$ remained small for one and two dimensional lattices.

A theoretical description of quantum depletion can be derived from the Bogoliubov theory which is the standard theory to describe the ground state properties of a weakly interacting system. The population in the (quasi-) momentum state $k$ is given by \cite{bog, stoof, rey}:
\begin{equation}
v_k^2=\frac{T(k)+n_0 U-\sqrt{2\, T(k) n_0 U+T^2(k)}}{2 \sqrt{2\, T(k) n_0 U+T^2(k)}}
\label{eq4}
\end{equation}
where $T(k)$ is the kinetic energy, $n_0$ is the occupancy number [per cubic lattice cell of $(\lambda_{\mathrm{latt}}/2)^3$] in the zero momentum state, and $U$ is the on-site interaction energy \cite{fisher,jaksch, krauth,freericks}. Incidentally, $v_k$ is one of the coefficients in the Bogoliubov transformation. The total occupancy number $n$ is given by the sum of $n_0$ and the quantum depletion: $n=n_0+\sum_{k\neq0}v_k^2$. For a given density $n$, the quantum depletion can be obtained by using Eq.~(\ref{eq4}) and the appropriate dispersion relation $T(k)$.

\begin{figure}[htbp]
\includegraphics[width=80mm]{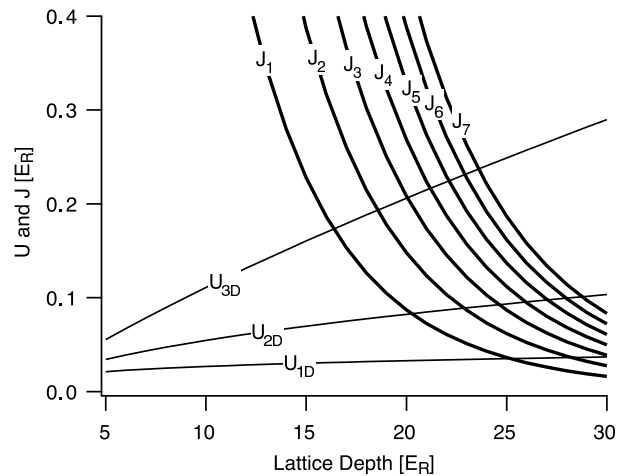}
\caption{On-site interaction $U$ and tunnelling rate $J$ for a $^{23}$Na BEC in an optical lattice at 1064~nm: $U_{d\mathrm{D}}$ is for a $d$-dimensional lattice. In a three dimensional lattice, the superfluid to Mott-insulator transition for occupancy number $n$ occurs when $J_n = 6 (2n+1+2\sqrt{n(n+1)}) J$ equals $U_{\mathrm{3D}}$ [see Eq.~(\ref{eq3})]. The horizontal locations of the crossing points where $J_n=U_{\mathrm{3D}}$ are the critical lattice depths.} \label{uandj}
\end{figure}

A band structure calculation was performed to obtain the on-site interaction $U$ (and also the tunnelling rate $J$) as functions of the lattice depths, shown in Fig.~\ref{uandj}. In calculating $U$, we use a Wannier density function along the dimensions with a lattice, and a uniform density in the ones without. $J$ is independent of the lattice wavelength or atomic mass for a given lattice depth (all energies measured in $E_R$).

The quantum depletion for a lattice of uniform occupation is obtained by integrating over the first Brillouin zone: $n=n_0+\int \,v_\textbf{k}^2\,d\textbf{k}$. For a sufficiently deep lattice ($\gtrsim 5~E_R$), the dispersion relation is given by:
\begin{equation}
T(\textbf{q})=4 J\,\sum_{i=1}^d\sin^2(q_{i}\,\pi)+4E_R \sum_{i=d}^3 q_i^2.
\label{eq7}
\end{equation}
where dimensions 1 through $d$ are assumed to have a lattice present and $\textbf{q}=\textbf{k}\,\lambda_{\mathrm{latt}}/4\pi$.

For an inhomogeneous system such as a harmonically confined condensate, we apply the result from the homogeneous system to shells of different occupancy numbers using the local density approximation (as the dependence of the quantum depletion on the occupancy number is slowly-varying). The calculated quantum depleted fractions are plotted in Fig.~\ref{123d}. The semi-quantitative agreement between the observed quantum depletion and the simple Bogoliubov theory, even for values around 50~\%, is the main result of this Letter. The remaining discrepancies may be due to unaccounted heating, a systematic overestimate of the background, and the inadequacies of Bogoliubov theory for large quantum depletion.

In free space, the dispersion curve is a continuous parabola. Both the number of populated states and the population in each state increases with the atomic density $\rho$, and the quantum depletion $\eta_0$ is proportional to $\rho^{1/2}$. This still holds for shallow lattices ($\lesssim 5~E_R$) when the quantum depletion ($\eta$) does not saturate the lowest band: $\eta=\eta_0 \left(\frac{\epsilon_{\mathrm{MF}}^{\ast} m^{\ast}}{\epsilon_{\mathrm{MF}}m}\right)^{3/2}$ ($\epsilon_{\mathrm{MF}}$ and $m$ are the free space mean-field energy and atomic mass; $\epsilon_{\mathrm{MF}}^{\ast}$ and $m^{\ast}$ are the lattice-enhanced mean-field energy and effective mass) \cite{pert}.

The situation changes for deeper lattices as the interaction energy $U$ becomes comparable to the width of the first energy band (approximately $4 J$). In this regime, the quantum depletion starts to saturate the lowest band, but the higher bands remain virtually empty due to the large band gap. While the population $N_{\mathrm{qd}}$ in the lowest band continues to increase with the atomic density $\rho=n/(\lambda_{\mathrm{latt}}/2)^3$, the quantum depleted fraction $N_{\mathrm{qd}}/N$ actually decreases with $\rho$ [see Fig.~\ref{123d} where the calculated quantum depletion is bigger for $n=1$ than for $n=7$ at large lattice depths ($\gtrsim 9~E_R$)].

The fact that the observed quantum depletion for one or two dimensional lattices remained small provides further evidence for our interpretation of the diffuse background as quantum depletion. In the dimension with a lattice present, the band width is proportional to the tunneling rate $J$ which decreases exponentially with the lattice depth. The interaction energy $U$ increases much slower with the lattice depth. Therefore the flattening of the dispersion relation contributes more significantly to the increased quantum depletion. Since this flattening does not occur in the dimension without a lattice, the quantum depletion for one or two dimensional lattices is expected to increase much slower compared to three dimensional lattice, consistent with our observation.

For a three dimensional lattice, quantum depletion and the superfluid to Mott-insulator transitions are two consequences of admixing higher momenta into the many-body wave function. For increasing interactions, the ground state wave function is first a depleted superfluid, then developes strong correlations and suppressed density fluctuations and finally turns into the Mott insulator.  In our measurements, the insulating regions appear fully in the diffuse background together with the quantum depletion of the superfluid regions. A lower bound for our measured $N_{\mathrm{qd}}/N$ is thereofre  provided by calculating the fraction of atoms in the Mott-insulator phase. According to the mean-field theory in a homogenous bosonic lattice system \cite{fisher,jaksch, krauth,freericks}, the critical value $(U/J)_c$ at which the phase transition occurs for occupancy number $n$ is given by:
\begin{equation}
(U/J)_c = z[2n+1+2\sqrt{n(n+1)})]\label{eq3}
\end{equation}
where $z$ is the number of nearest neighbors ($z=2d$ for a $d$-dimensional cubic lattice). For an inhomogeneous system such as a trapped condensate, shells of different occupancy numbers enter the Mott-insulator phase from outside as the lattice potential is increased and $U/J$ exceeds the critical values.

In our experiment, the peak occupancy number is $\sim7$. From Fig.~\ref{uandj}, the critical lattice depths in a three dimensional cubic lattice for $n=1,2,...,7$ are 16.3, 18.5, 20.0, 21.2, 22.1, 22.9, 23.6~$E_{R}$ respectively. The integrated Mott-insulator fraction $N_{\mathrm{MI}}/N$ as a function of lattice depths is plotted in Fig.~\ref{123d}. Instead of a step function with jumps at each critical lattice depth, we use a smooth spline curve for $N_{\mathrm{MI}}/N$, which is more realistic given the fluctuations associated with the atom numbers and lattice depths. The measured $N_{\mathrm{qd}}/N$ was significantly greater than $N_{\mathrm{MI}}/N$.

In the case of one and two dimensional lattices, a Mott-insulator transition would only occur for lattice depths much larger than those in our experiment. Note that Eq.~(\ref{eq3}) is not directly applicable as the dimensions without lattice beams in our system are only loosely confined and cannot be considered frozen\cite{krauth}. In addition, $n$ in Eq.~(\ref{eq3}) is the number of atoms \emph{per lattice site} and far exceeds 7 for the lower-dimensional lattices.

In conclusion, we conducted a quantitative study of quantum depletion in a gaseous BEC through the application of an optical lattice, and found reasonable agreement with a model based on Bogoliubov theory in the predominantly superfluid regime. A complementary study was recently reported by Gerbier $\emph{et al..}$ \cite{gerbier1,gerbier2} of the non-vanishing visibility of the interference peaks in a Mott insulator as a result of the admixture of particle-hole states. The two works together give a complete description of the ground state in both the superfluid and insulating phases. More elaborate theoretical treatments for the intermediate case have been presented in Refs \cite{seng05, garc04, schr04}.

The authors would like to thank Jamil Abo-Shaeer and Takashi Mukaiyama for experimental assistances. This research is supported by NSF, ONR, ARO, and NASA.

\bibliographystyle{apsrev}
\bibliography{qd}

\begin{thebibliography}{32}
\bibitem[{per()}]{pert}
\bibinfo{note}{For very shallow lattices, one can obtain from perturbation
  theory: $m^{\ast}/m=1+E_{\mathrm{latt}}^2/32$,
  $\epsilon_{\mathrm{MF}}^{\ast}/\epsilon_{\mathrm{MF}}=1+3E_{\mathrm{latt}}^2%
/32$, where $E_{\mathrm{latt}}$ is the lattice depth divided by $E_R$.}
\end{thebibliography}

\end{document}